\documentclass[submission,copyright,creativecommons]{eptcs}
\providecommand{\event}{GASCOM 2024} % Name of the event you are submitting to

\usepackage{iftex}

\ifpdf
  \usepackage{underscore}         % Only needed if you use pdflatex.
  \usepackage[T1]{fontenc}        % Recommended with pdflatex
\else
  \usepackage{breakurl}           % Not needed if you use pdflatex only.
\fi

\usepackage{graphicx} % Required for inserting images
\usepackage[english]{babel}
\usepackage{amsthm}
\usepackage[normalem]{ulem}
\usepackage{amssymb}
\usepackage[toc,page]{appendix}
\usepackage{algorithm}
\usepackage{amsmath}
\usepackage[noend]{algpseudocode}
\usepackage{enumitem}
\usepackage{xcolor}
\usepackage{multicol}
\makeatletter
\def\BState{\State\hskip-\ALG@thistlm}
\makeatother
\newtheorem{theorem}{Theorem}[section]

\newtheorem{proposition}[theorem]{Proposition}
\newtheorem{lemma}[theorem]{Lemma}

\theoremstyle{definition}
\newtheorem{definition}[theorem]{Definition} 
 
\newtheorem{rk}[theorem]{Remark}
\newtheorem{fact}[theorem]{Fact}

\def\N{\mathbb{N}}

\def\R{\mathbb{R}}
\def\D{\mathcal{D}}
\def\F{\mathcal{F}}

\title{Greedy Gray Codes for some Restricted \\ Classes of Binary Words}
\author{Nathanaël Hassler \qquad\qquad Vincent Vajnovszki
\institute{Laboratoire d'Informatique de Bourgogne\\ Dijon, France}
\email{\quad nathanael.hassler@ens-rennes.fr \quad\qquad vincent.vajnovszki@u-bourgogne.fr}
\and 
Dennis Wong
\institute{Macao Polytechnic University\\Macao, China}
\email{cwong@uoguelph.ca}}
\def\titlerunning{Greedy Gray Codes for some Restricted Classes of Binary Words}
\def\authorrunning{N. Hassler, V. Vajnovszki, D. Wong}
\begin{document}
\maketitle

\begin{abstract}
We investigate the existence of greedy Gray codes, based on the choice of the first element in the code, for two classes of binary words: generalized Fibonacci words and generalized Dyck words.
\end{abstract}

\section{Introduction}

\subsection{Constrained binary words}

\subsubsection*{1's run constrained binary words}
\noindent

Let $n,p\in \mathbb{N}$, $p\geq2$, and $F_n(p)$ be the set of length $n$ binary words with no $p$ consecutive 1's. $F_n(2)$ is counted by the Fibonacci numbers $f_n$, and in general, $F_n(p)$ is counted by the $p$-order Fibonacci numbers $f_n^{(p)}$.
Now let $k\in \mathbb{N}$, and $F_n(p,k)$ be the subset of words in $F_n(p)$ of weight $k$
(i.e., with exactly $k$ 1's). $F_n(p,k)$ is counted by the univariate $p$-nomial coefficient.

\subsubsection*{Prefix constrained binary words}
\label{prefix_constr}
Let $k,n,p\in \mathbb{N}$ with $(p+1)k\leq n$, and
$C_n(p,k)$ be the set of length $n$ binary words of weight $k$ with the property that any prefix contains at least $p$ times as many $0$'s as 1's. In particular:
\begin{itemize}
\item $C_n(0,k)$ is the set of length $n$ binary words of weight $k$ (combinations in binary word representation),
\item $C_{2n}(1,n)$ is the set of length $2n$ Dyck words, and it is counted by the Catalan numbers $\binom{2n}{n}-\binom{2n}{n-1}=\frac{1}{n+1}\binom{2n}{n}$,
\item $C_{3n}(2,n)$ is in bijection with size $3n$ ternary trees
(see A001764 in \cite{OEIS}).
\end{itemize}

More generally, $C_{(p+1)n}(p,n)$ is counted by $\frac{1}{pn+1}\binom{(p+1)n}{n}$, known as the Pfaff–Fuss–Catalan numbers, and the cardinality of $C_n(p,k)$ is established for instance in \cite[Equation (2)]{Vajnovszki-Walsh}, using generating functions:
\begin{equation}
    |C_n(p,k)|=\binom{n}{k}-p\binom{n}{k-1}.
\end{equation}

\subsection{Gray codes and greedy algorithms}
A Gray code for a class of combinatorial objects is a list that contains each object from the class exactly once such that any two consecutive objects in the list differ only by a `small change' \cite{Mutze}. In this paper we restrict ourselves to Gray codes for restricted classes of same length and same weight binary words
(the weight of a binary word being its number of $1$'s). The `small changes' we consider here are  homogeneous transpositions: two binary words differ by a {\em homogeneous} transposition if
one can be obtained from the other
by transposing a $1$ with a $0$, and there are no $1$'s between the transposed bits. A Gray code is called homogeneous if consecutive words differ in a such a way.
A list of words is {\em suffix partitioned} if words with the same suffix are consecutive in the list.

The next definition of the {\em greedy Gray code algorithm} is a specialisation of that introduced in \cite{Williams} to particular cases of binary words, see also \cite{Wong-Vajnovszki}.

\begin{definition}
\label{def_greedy}
For a set $S$ of same length and same weight binary words the {\em greedy Gray code algorithm} to obtain a Gray code list $\mathcal{L}$ for the set $S$ is:
\begin{enumerate}
\item Initialize $\mathcal{L}$ with a particular word in $S$. 
\item For the last word in $\mathcal{L}$, homogeneously transposes the leftmost possible 1 with the leftmost possible $0$, such that the obtained word is in $S$ but not in $\mathcal{L}$.
\item If at point 2. a new word is obtained, then append it to the list $\mathcal{L}$ and return to point 2. 
\end{enumerate}
\end{definition}

In the following we will say that the list $\mathcal{L}$  is obtained by applying the greedy algorithm for $S$ to $\alpha$, where $\alpha$ is the initial word of $\mathcal{L}$.
Depending on the choice of $\alpha$, it can happen that the obtained list $\mathcal{L}$ is not an exhaustive one for~$S$.

\section{Tail partitioned lists}

The tail of a binary word is its unique suffix of the form $011\cdots 1$, and the only words with no tail have the form $11\cdots 1$. Note that $0$ is also a tail, for the words ending by $0$.

A list of binary words is {\em increasing} ({\em decreasing}) {\em tail partitioned} if words with tails of length $\ell$ appear before (after) words with tail of length $\ell+1$, for any $\ell \geq 1$.

\begin{definition}
A list $\mathcal L$ of same length binary words is {\it recursive tail partitioned}~if it is empty, or 
\begin{itemize}
\item it is increasing or decreasing tail partitioned, and 
\item for any tail $t$, the list obtained by: (i) considering the sublist of $\mathcal L$ of words with tail $t$, then (ii) erasing the tail $t$ in each word of this sublist, is in turn recursive tail partitioned.
\end{itemize}
    
\end{definition}

In the following $\cdot$ denotes the concatenation (of two words, or of each word in a list with a word) and the comma appends lists. With this notation, $\mathcal L$ is a recursive tail partitioned list  if it is empty or has the form 

\begin{equation}
\label{tp1}
\mathcal{L}  =\mathcal{L}_1\cdot 01^u, \mathcal{L}_2\cdot  01^{u+1}, \mathcal{L}_3\cdot  01^{u+2}, \cdots, \mathcal{L}_{\ell+1}\cdot  01^{u+\ell} 
\end{equation}
or the form
\begin{equation}
\label{tp2}
\mathcal{L}  = \mathcal{L}_1\cdot 01^{u+\ell}, \mathcal{L}_2\cdot  01^{u+\ell-1}, \mathcal{L}_3\cdot  01^{u+\ell-2}, \cdots, \mathcal{L}_{\ell +1}\cdot  01^u \\
\end{equation}
for some $u,\ell\geq 0$,
and each list $\mathcal{L}_i$, is in turn recursive tail partitioned.

We remark that the list in (\ref{tp1}) is not necessarily the reverse of that in (\ref{tp2}).
Clearly, a recursive tail partitioned list (r-t partitioned list for short) is a suffix partitioned list.

\begin{theorem}\label{main}
    If the list $\mathcal{L}$ is a homogeneous and suffix partitioned Gray code for a set of (same length and same weight) binary words, then $\mathcal{L}$ is an r-t partitioned list.
\end{theorem}

\section{Main results}

\subsection{Fibonacci words $F_n(2,k)$}

In this subsection we show that for any $\alpha\in F_n(2,k)$, by applying the greedy algorithm for $F_n(2,k)$ to $\alpha$, a suffix partitioned list is obtained, and we characterize the words $\alpha$ such that the greedy algorithm yields an exhaustive list for $F_n(2,k)$.
Moreover, for every $\alpha\in F_n(2,k)$ we characterize the last word in the obtained list.

\medskip
For $n<2k-1$, $F_n(2,k)$ is empty and 
in two particular cases $F_n(2,k)$ is a singleton set.
\begin{fact}
\label{small_r}
If $k=0$, then $F_n(2,k)=\{0^n\}$, and if  $n=2k-1$, then $F_n(2,k)=\{1(01)^{k-1}\}$.
\end{fact}

For $\alpha\in F_n(2,k)$ we denote by $\F(\alpha)$ the list obtained by applying the greedy algorithm for $F_n(2,k)$ to $\alpha$.

For $n,k$ with $n\geq 2k$, let $\alpha_{n,k}^i=0^i1(01)^{k-1}0^{n-2k+1-i}$, for $0\leq i\leq n-2k+1$. Furthermore, let $\gamma_{n,k}:=\alpha_{n,k}^{n-2k+1}=0^{n-2k}(01)^k$. %Clearly, when $k=0$, then $\alpha_{n,k}=\beta_{n,k}=\gamma_{n,k}=0^n$, and when $n=2k$, then $\alpha_{n,k}=\beta_{n,k}=(10)^k$. 
We denote by $\textit{GenF}(n,k)$ the set of words $\alpha\in F_n(2,k)$ such that $\F(\alpha)$ is a homogeneous Gray code for $F_n(2,k)$. Equivalently, $\alpha\in\textit{GenF}(n,k)$ if and only if $\F(\alpha)$ contains every word of $F_n(2,k)$.

\begin{theorem}\label{greedy fib}
    Let $n\in\N^\star$. Then for all $k$ with $n\geq 2k$, we have 
    \begin{enumerate}
        \item $\alpha_{n,k}^i\in\textit{GenF}(n,k)$ for all $0\leq i\leq n-2k$, and $\F(\alpha_{n,k}^i)$ is a suffix partitioned list, with $\gamma_{n,k}$ as last word.
        \item $\gamma_{n,k}\in\textit{GenF}(n,k)$, $\F(\gamma_{n,k})$ is a suffix partitioned list and its last word is
        \begin{enumerate}[label=(\roman*)]
            \item $\alpha_{n,k}^0$ if $k$ is even,
            \item $\alpha_{n,k}^{n-2k}$ if $k$ is odd.
        \end{enumerate}
    \end{enumerate}
\end{theorem}

In the next lemma, we extend the result of 1. in Theorem \ref{greedy fib} concerning the last element of $\F(\alpha)$ to each $\alpha\in F_n(2,k)$.

\begin{lemma}\label{last word F(alpha)}
    Let $n\geq 2k$. For any $\alpha\in F_n(2,k)$ with $\alpha\ne \gamma_{n,k}$, the last word of $\F(\alpha)$ is $\gamma_{n,k}$.
\end{lemma}

Now we generalise the property for $\F(\alpha)$ to be suffix partitioned to each word of $F_n(2,n)$.

\begin{lemma}\label{F(alpha) is suffix partitioned}
    Let $n\geq 2k-1$. For any $\alpha\in F_n(2,k)$, $\F(\alpha)$ is a suffix partitioned list. 
\end{lemma}

Now we are able to completely describe the set $\textit{GenF}(n,k)$.

\begin{proposition}\label{generators F_n(2,k)}
    Let $n\geq 2k-1$. Then 
    $$\textit{GenF}(n,k)=\{0^i1(01)^{k-1}0^{n-2k+1-i} \ | \ 0\leq i \leq n-2k+1\}.$$
    In particular, $|\textit{GenF}(n,k)|=n-2k+2$.
\end{proposition}

\subsection{$C_n(p,k)$}

For $n,p,k$ such that $n\geq (p+1)k$ and $\alpha\in C_n(p,k)$, we denote by $\D_p(\alpha)$ the list obtained by applying the greedy algorithm for $C_n(p,k)$ to $\alpha$. Let also $\textit{Gen}_p(n,k)$ be the set of words $\alpha\in C_n(p,k)$ such that $\D_p(\alpha)$ is a homogeneous Gray code for $C_n(p,k)$. Equivalently, $\alpha\in\textit{Gen}_p(n,k)$ if and only if $\D_p(\alpha)$ contains every word of $C_n(p,k)$. Since the situation is a bit more complicated than the one for Fibonacci words, for clarity we first investigate the case $p\in\N$, explaining every details of the proofs. Then we explain how the results are generalised to any $p\in\R$, not necessarily describing each proof.

\subsubsection{$p\in\N$}

We start by investigating the case $p\in\N$. We fix such a $p$ throughout this section. For $n\geq (p+1)k$, let $\alpha_{n,k}^{i,j}:=0^{pj-i}1^{j-1}0^i1(0^p1)^{k-j}0^{n-(p+1)k}$ (resp. $\alpha_{n,k}=1^k0^{n-k}$) for $i=0,\ldots,p-1$ and $j=1,\ldots,k$ if $p\geq 1$ (resp. if $p=0$), and $\beta_{n,k}^i:=0^i1^k0^{n-i-k}$ for $i=pk+1,\ldots,n-k$. Because it plays a special role, we will use different notation for $\beta_{n,k}^{n-k}$, so we set $\gamma_{n,k}:=\beta_{n,k}^{n-k}=0^{n-k}1^k$. %Theorem \ref{main greedy C_n(1,k)} can then be generalised as follows.

\begin{theorem}\label{main greedy C_n(p,k)}
    Let $n\geq (p+1)k$, with $k\geq 1$.
    \begin{enumerate}
        \item For all $0\leq i\leq p-1$ and $1\leq j\leq k$, we have $\alpha_{n,k}^{i,j}\in\textit{Gen}_p(n,k)$ (resp. $\alpha_{n,k}\in\textit{Gen}_0(n,k)$), and $\D_p(\alpha_{n,k}^{i,j})$ (resp. $\D_0(\alpha_{n,k})$) is a suffix partitioned list with $\gamma_{n,k}$ as last word.
        \item For all $j$, $pk+1\leq j\leq n-k-1$, we have $\beta_{n,k}^j\in\textit{Gen}_p(n,k)$, and $\D_p(\beta_{n,k}^j)$ is a suffix partitioned list with $\gamma_{n,k}$ as last word.
        \item $\gamma_{n,k}\in\textit{Gen}_p(n,k)$, $\D_p(\gamma_{n,k})$ is a suffix partitioned list and its last word is 
        \begin{enumerate}[label=(\roman*)]
            \item $0^p1$ if $(n,k)=(p+1,1)$,
            \item $\beta_{n,k}^{n-k-1}=0^{n-k-1}1^k0$ if $k$ is odd and $n\ne (p+1)k$,
            \item $\alpha_{n,k}^{1,k-1}=0^{(k-1)p-1}1^{k-2}010^p1$ if $k\geq 3$ is odd and $n=(p+1)k$ (resp. $\alpha_{n,k}=1^k$ if $p=0$),
            \item $\alpha_{n,k}^{1,k}=0^{kp-1}1^{k-1}010^{n-(p+1)k}$ if $k$ is even (resp. $\alpha_{n,k}=1^k0^{n-k}$ if $p=0$).
        \end{enumerate}
    \end{enumerate}
\end{theorem}

\begin{lemma}\label{last word of D_p(alpha)}
    Let $n\geq (p+1)k$. For all $\alpha\in C_n(p,k)$, with $\alpha\ne \gamma_{n,k}$, $\D_p(\alpha)$ ends with $\gamma_{n,k}$.
\end{lemma}

\begin{lemma}\label{D_p(alpha) is suffix partitioned}
    Let $n\geq 1$. Then for all $k$ with $n\geq (p+1)k$, and for each $\alpha\in C_n(p,k)$, $\D_p(\alpha)$ is a suffix partitioned list.
\end{lemma}

\begin{proposition}\label{generators C_n(p,k)}
    Let $n\geq (p+1)k$. If $p\geq1$ then 
    \begin{align*}
        \textit{Gen}_p(n,k)= \bigcup_{j=1}^k\{0^{pj-i}&1^{j-1}0^i1(0^p1)^{k-j}0^{n-(p+1)k} \ | \ 0\leq i\leq p-1\}\\
        &\cup\{0^i1^k0^{n-i-k} \ | \ pk+1\leq i\leq n-k\}.
    \end{align*}
    If $p=0$ then $\textit{Gen}_0(n,k)=\{0^i1^k0^{n-i-k} \ | \ 0\leq i\leq n-k\}$.
    In particular, $$|\textit{Gen}_p(n,k)|=
    n-k+1-p.$$
\end{proposition}

\begin{rk}
    Propositions \ref{generators F_n(2,k)} and \ref{generators C_n(p,k)} highlight the fact that the choice of the first element for the greedy algorithm is crucial. Indeed, only a few elements will produce a Gray code for $F_n(2,k)$ or $C_n(p,k)$ with this algorithm.
\end{rk}

\subsubsection{$p\in \R_+$}

The previous results can be generalized to $C_n(p,k)$ for any $p\in\R_+$. In particular we have the following result:

\begin{theorem}
    Let $p\in\R_+$ and $n\geq (p+1)k$. Then 
    $$|\textit{Gen}_p(n,k)|=n-k+1-\lceil p\rceil.$$
\end{theorem}

\subsection{Algorithmic considerations}

 Building on the previous results, the last part presents CAT algorithms that greedily generate Gray code for $C_n(p, k)$ and $F_n(2,k)$, which we omit in this abstract.
 
\nocite{*}
\bibliographystyle{eptcs}
\bibliography{generic}

\begin{thebibliography}{1}
\providecommand{\bibitemdeclare}[2]{}
\providecommand{\surnamestart}{}
\providecommand{\surnameend}{}
\providecommand{\urlprefix}{Available at }
\providecommand{\url}[1]{\texttt{#1}}
\providecommand{\href}[2]{\texttt{#2}}
\providecommand{\urlalt}[2]{\href{#1}{#2}}
\providecommand{\doi}[1]{doi:\urlalt{https://doi.org/#1}{#1}}
\providecommand{\eprint}[1]{arXiv:\urlalt{https://arxiv.org/abs/#1}{#1}}
\providecommand{\bibinfo}[2]{#2}

\bibitemdeclare{article}{Bultena-Ruskey}
\bibitem{Bultena-Ruskey}
\bibinfo{author}{A.~\surnamestart Bultena\surnameend} \&
  \bibinfo{author}{F.~\surnamestart Ruskey\surnameend} (\bibinfo{year}{1998}):
  \emph{\bibinfo{title}{An Eades-McKay algorithm for well-formed parentheses
  strings}}.
\newblock {\slshape \bibinfo{journal}{Information Processing Letters}}
  \bibinfo{volume}{68}, pp. \bibinfo{pages}{255--259},
  \doi{10.1016/S0020-0190(98)00171-9}.

\bibitemdeclare{article}{Eades-McKay}
\bibitem{Eades-McKay}
\bibinfo{author}{P.~\surnamestart Eades\surnameend} \&
  \bibinfo{author}{B.~\surnamestart McKay\surnameend} (\bibinfo{year}{1984}):
  \emph{\bibinfo{title}{An algorithm for generating subsets of fixed size with
  a strong minimal change property}}.
\newblock {\slshape \bibinfo{journal}{Information Processing Letters}}
  \bibinfo{volume}{19}, pp. \bibinfo{pages}{131--133},
  \doi{10.1016/0020-0190(84)90091-7}.

\bibitemdeclare{article}{Mutze}
\bibitem{Mutze}
\bibinfo{author}{Torsten \surnamestart Mütze\surnameend}
  (\bibinfo{year}{2023}): \emph{\bibinfo{title}{Combinatorial Gray codes -- an
  updated survey}}.
\newblock {\slshape \bibinfo{journal}{Electronic Journal of Combinatorics}}
  \bibinfo{volume}{30}(\bibinfo{number}{3}):\bibinfo{eid}{DS26},
  \doi{10.37236/11023}.

\bibitemdeclare{misc}{OEIS}
\bibitem{OEIS}
\bibinfo{author}{N.J.A. \surnamestart Sloane\surnameend}:
  \emph{\bibinfo{title}{The On-Line Encyclopedia of Integer Sequences}}.
\newblock \urlprefix\url{http://oeis.org/}.

\bibitemdeclare{article}{Vajnovszki-Walsh}
\bibitem{Vajnovszki-Walsh}
\bibinfo{author}{V.~\surnamestart Vajnovszki\surnameend} \&
  \bibinfo{author}{T.~\surnamestart Walsh\surnameend} (\bibinfo{year}{2006}):
  \emph{\bibinfo{title}{A loop-free two-close Gray code algorithm for listing
  $k$-ary Dyck words}}.
\newblock {\slshape \bibinfo{journal}{Journal of Discrete Algorithms}}
  \bibinfo{volume}{4}(\bibinfo{number}{4}), pp. \bibinfo{pages}{633--648},
  \doi{10.1016/j.jda.2005.07.003}.

\bibitemdeclare{inproceedings}{Williams}
\bibitem{Williams}
\bibinfo{author}{A.~\surnamestart Williams\surnameend} (\bibinfo{year}{2013}):
  \emph{\bibinfo{title}{The greedy Gray code algorithm}}.
\newblock In: {\slshape \bibinfo{booktitle}{Proceedings of the 13th
  international conference on Algorithms and Data Structures}}, p.
  \bibinfo{pages}{525–536},
  \doi{10.1007/978-3-642-40104-6_46}.

\bibitemdeclare{inproceedings}{Wong-Vajnovszki}
\bibitem{Wong-Vajnovszki}
\bibinfo{author}{D.~\surnamestart Wong\surnameend} \&
  \bibinfo{author}{V.~\surnamestart Vajnovszki\surnameend}
  (\bibinfo{year}{2023}): \emph{\bibinfo{title}{Greedy Gray codes for Dyck
  words and ballot sequences}}.
\newblock In: {\slshape \bibinfo{booktitle}{Computing and Combinatorics. COCOON
  2023}}, {\slshape \bibinfo{series}{Lecture Notes in Computer Science}}
  \bibinfo{volume}{14423}, \bibinfo{publisher}{Springer}, pp.
  \bibinfo{pages}{29--40}, \doi{10.1007/978-3-031-49193-1_3}.

\end{thebibliography}
\end{document}